\documentclass{aastex}

\def\arcsecpoint{$''\!.$}

\received{}
\accepted{}

\slugcomment{Submitted to the {\it The Astrophysical Journal (Letters)}}

\shorttitle{Intrinsic Absorption in NGC~4151}
\shortauthors{Crenshaw et al.}

\begin{document}

\title{STIS Echelle Observations of NGC 4151: Variable Ionization 
of the Intrinsic UV Absorbers\altaffilmark{1}}

\author{D.M. Crenshaw\altaffilmark{2,3},
S.B. Kraemer\altaffilmark{2},
J.B. Hutchings\altaffilmark{4},
A.C. Danks\altaffilmark{5},
T.R. Gull\altaffilmark{6},\\
M.E. Kaiser\altaffilmark{7},
C.H. Nelson\altaffilmark{8},
\& D. Weistrop\altaffilmark{8}
}

\altaffiltext{1}{Based on observations made with the NASA/ESA Hubble Space 
Telescope. STScI is operated by the Association of Universities for Research in 
Astronomy, Inc. under NASA contract NAS5-26555. }

\altaffiltext{2}{Catholic University of America and Laboratory for Astronomy and 
Solar Physics, NASA's Goddard Space Flight Center, Code 681,
Greenbelt, MD  20771}

\altaffiltext{3}{crenshaw@buckeye.gsfc.nasa.gov}

\altaffiltext{4}{Dominion Astrophysical Observatory, National Research
Council of Canada, 5071 W. Saanich Rd., Victoria, B.C. V8X 4M6, Canada}

\altaffiltext{5}{Raytheon Polar Services (RPSC), NASA's Goddard Space Flight 
Center, Code 681, Greenbelt, MD  20771}

\altaffiltext{6}{NASA's Goddard Space Flight Center, Laboratory for Astronomy
and Solar Physics, Code 681, Greenbelt, MD 20771}

\altaffiltext{7}{Department of Physics and Astronomy, Johns Hopkins University,
Baltimore, MD 21218}

\altaffiltext{8}{Department of Physics, University of Nevada, Las Vegas,
4505 Maryland Parkway, Las Vegas, NV 89154-4002}

\begin{abstract}

We present echelle observations of the intrinsic UV absorption lines in the 
Seyfert galaxy NGC~4151, which were obtained with the Space Telescope Imaging 
Spectrograph (STIS) on the Hubble Space Telescope ({\it HST}) on 1999 July 19. 
The UV continuum flux at 1450 \AA\ decreased by factor of about four over the 
previous two years and there was a corresponding dramatic increase in the column 
densities of the low-ionization absorption lines (e.g., Si~II, Fe~II, and 
Al~II), presumably as a result of a decrease in the ionizing continuum.
In addition to the absorption lines seen in previous low states, we identify a 
large number of Fe~II absorption lines that arise from metastable levels as high 
as 4.1 eV above the ground state, indicating high densities ($>$ 10$^{6}$ 
cm$^{-3}$). We find that the transient absorption feature in the blue wing of 
the broad C~IV emission, seen in a Goddard High Resolution Spectrograph ({\it 
GHRS}) spectrum and thought to be a high-velocity C~IV component, is actually a 
Si~II fine-structure absorption line at a radial velocity of $-$560 km s$^{-1}$ 
(relative to systemic). We also demonstrate that the ``satellite''
emission lines of C~IV found in International Ultraviolet Explorer ({\it IUE}) 
spectra are actually regions of unabsorbed continuum plus broad emission that 
become prominent when the UV continuum of NGC~4151 is in a low state.

\end{abstract}

\keywords{galaxies: individual (NGC 4151) -- galaxies: Seyfert}

~~~~~

\newpage

\section{Introduction}

NGC~4151 (cz $=$ 995 km s$^{-1}$) is the first Seyfert galaxy known to show
intrinsic absorption that could be attributed to the active nucleus. Oke and 
Sargent (1968) first reported evidence for nonstellar absorption from He~I 
$\lambda$3889, and Anderson \& Kraft (1969) discovered H$\beta$ and H$\gamma$ 
self absorption.
The optical absorption was blueshifted with radial velocities up to $-$970 km 
s$^{-1}$ with respect to the host galaxy and therefore attributed to mass 
ejected from the nucleus. Cromwell \& Weymann  (1970) discovered that the Balmer 
absorption is variable, and can sometimes disappear altogether.

Ultraviolet observations of NGC~4151 by the {\it IUE} (Boksenberg et al. 
1978) and subsequent far-ultraviolet observations by the {\it Hopkins 
Ultraviolet Telescope} ({\it HUT}, Kriss et al. 1992) revealed a number of 
absorption lines from species that span a wide range in ionization potential 
(e.g., O~I to O~VI), as well as fine-structure and metastable absorption lines.
The intrinsic UV absorption was found to be variable in ionic column 
density, but no variations in radial velocities were detected (Bromage et 
al. 1985). More recent observations of NGC 4151 were obtained with the Goddard 
High Resolution Spectrograph (GHRS) at high spectral resolution ($\sim$15 km 
s$^{-1}$) over limited wavelength regions by Weymann et al. (1997). The GHRS 
spectra revealed that the 
C~IV and Mg~II absorption lines, detected in six major kinematic 
components, were remarkably stable over the time period 1992 -- 1996.
The absorption components showed no evidence for variable column or radial 
velocity, except that a transient feature was detected in the blue wing of the 
broad C~IV emission in one of four GHRS observations, and was attributed to a 
high-velocity C~IV component (Weymann et al. 1997).

In the Hubble Space Telescope ({\it HST}) survey of intrinsic UV absorption 
lines in Seyfert galaxies by Crenshaw et al. (1999), NGC~4151 stood out as an 
unusual object. Of the ten Seyfert galaxies in this sample with intrinsic 
absorption, all showed C~IV and N~V absorption, but only NGC 4151 
showed Mg~II absorption
\footnote{Interestingly, a similar percentage ($\sim$15\%) of broad 
absorption-line (BAL) QSOs show low-ionization absorption in the form of Mg~II 
(Weymann et al. 1991).}.
NGC 4151 is also unusual in the sense that it is the only active galaxy 
known to show metastable C~III$^{*}$ $\lambda$1175 absorption, which is 
indicative of relatively high electron densities ($\sim$ 10$^{9}$ cm$^{-3}$, 
Bromage et al. 1985).

Due to the unusual nature of the absorption in NGC~4151, we decided to obtain 
STIS echelle spectra of the UV spectrum from 1150 -- 3100 \AA\ at a spectral 
resolution of 7 -- 10 km s$^{-1}$ to study the 
intrinsic absorption in detail. These observations are part of a long-term 
project on NGC~4151 by members of the Instrument Definition Team (IDT) of the 
Space Telescope Imaging Spectrograph (STIS) on {\it HST}. In this 
letter, we concentrate on the variability of the absorption features in the 
spectral region surrounding the broad C~IV emission. In future papers, we will 
present results on the entire UV spectrum, the Galactic absorption, and the 
kinematics and physical conditions in the intrinsic absorbers.

\section{Observations}

We observed the nucleus of NGC~4151 with the medium-resolution echelle gratings 
of STIS on 1999 July 19/20. Table 1 gives the details of the new observations, 
as well as previous STIS spectra of the nucleus in the UV that we use for 
comparison. The slitless G140M spectrum is discussed in Hutchings et al. (1998) 
and the low-resolution G140L and G230LB spectra are described by Nelson et al. 
(2000). We reduced the echelle spectra using the IDL software developed at 
NASA's Goddard Space Flight Center for the IDT. The data reduction included a 
procedure to remove the background light from each order using a scattered-light 
model devised by Lindler (1999) rather than the standard procedure, which 
estimates the background from the interorder light directly. Whereas the 
standard procedure resulted in significant negative fluxes (up to 10\% of the 
continuum) for the troughs of saturated interstellar lines (such as Ly$\alpha$ 
and Si~II $\lambda$1260.4), the improved procedure brought all of these troughs 
close to zero (i.e., to within 1\% of the continuum flux levels). For display 
purposes, the individual orders in each echelle spectrum were spliced together 
in the regions of overlap.

Figure 1 shows the C~IV region for the three epochs of STIS observations that 
cover this region. The major kinematic components of C~IV absorption identified 
by Weymann et al. (1997) in the GHRS spectra are labeled. These components have 
velocity centroids 
that range from $-$1575 to $+$38 km s$^{-1}$, relative to the systemic redshift 
of cz $=$ 995 km s$^{-1}$ (from H~I 21 cm observations, see de Vaucouleurs et 
al. 1991), and appear in all of the GHRS and STIS spectra. As noted by Weymann 
et al., component B originates in our galaxy and component F probably arises in 
the interstellar medium and/or halo of NGC~4151. There are no obvious changes in 
the velocity centroids or widths of the C~IV components since the GHRS 
observations, except for a possible new broad component in the 1540 -- 1547 \AA\ 
region (see Kraemer et al. 2000 for more details).

It is obvious from Figure 1 that the continuum plus broad-line emission 
decreased considerably over the two-year period of STIS observations. The 
continuum flux at 1450~\AA\ dropped by a factor of 2.3 between the last two 
observations  (i.e., from 2.45 x 10$^{-13}$ to 1.06 x 10$^{-13}$ ergs s$^{-1}$ 
cm$^{-2}$ \AA$^{-1}$ between 1998 February 10 and 1999 July 19).
There are no pure continuum regions in the slitless spectrum obtained on 1997 
May 25, but the continuum plus broad emission is 1.5 times higher than in 1998.
Since the amplitude of continuum variations tends to be larger than that of C~IV 
in this object (Crenshaw et al. 1996, and references therein), it is likely that 
the UV continuum at 1450 \AA\ decreased by at least a factor of 1.5 from 
1997 to 1998, and therefore by at least a factor of 3.5 from 1997 to 1999. This 
is not unusual for NGC~4151, since {\it IUE} observations have shown that the UV 
continuum can vary by as much as a factor of 10 over a period of 2 -- 3 years 
(Ulrich et al. 1991). 

\section{Results}

\subsection{The Appearance of Low-Ionization Lines}

Figure 1 shows an interesting change in the C~IV region at 
the time of the STIS echelle observations, when the continuum and broad-line 
fluxes are low. A number of new broad (FWHM $\approx$ 400 km s$^{-1}$) 
absorption features appear in the wings of C~IV which are not seen at higher 
states in the previous GHRS and STIS observations (with an exception discussed 
in Section 3.3); the strongest of these features are at the observed wavelengths 
of 1529 \AA, 1535 \AA, 1572 \AA, and 1577 \AA. These 
absorption features and others are identified in Figure 2, which gives a 
wider region of the spectrum near C~IV.
We confirm Bromage et al.'s (1985) identification of Si~IV, Si~II, C~IV, and 
Al~II absorption in this region, as well as their claim of resonance and 
fine-structure lines from multiplet UV8 of Fe II.
These broad features are primarily due to component D (at a radial velocity of 
$-$560 km s$^{-1}$ with respect to the host galaxy), although some 
lines receive a contribution from component E (at $-$255~km~s$^{-1}$; a full 
deconvolution will be given in Kraemer et al. 2000).

Many other absorption features are present in the spectrum in Figure 2. We 
attribute a majority of these features to Fe~II multiplets that arise from 
metastable levels up to 4.1 eV above the ground state (Silvis \& Bruhweiler 
2000). As shown in Figure 2, the correspondence of the absorption features with 
expected positions of the Fe II lines at the radial velocity of component D is 
convincing. Assuming collisional excitation of the metastable Fe~II levels, this 
implies densities of n$_{e}$ $>$ 10$^{6}$ cm$^{-3}$ in this kinematic component 
(Wampler, Chugai, \& Pettijean 1995). This is probably a 
severe lower limit, since densities on the order of 10$^{9}$ cm$^{-3}$ 
have been derived from the presence of metastable C~III$^{*}$ $\lambda$1176 
(from a level 6.5 eV above the ground state) at this 
approximate radial velocity (Espey et al. 1998). A number of lines in Figure 
2 remain unidentified, and we suspect that some of these may arise from higher 
Fe~II levels (not on the lists of Fe~II multiplets in Silvis and Bruhweiler). We 
note that the effect of all this absorption is to add structure to the UV 
spectrum and depress the apparent continuum and broad-line emission in many 
places. For example, the unusual looking feature between 1593 and 1602 \AA\ is 
an unabsorbed portion of the broad wing of C~IV, located between two Fe~II 
absorption lines from multiplets 44 and EO.

\subsection{Variable ionization of the absorption}

There are two likely sources of intrinsic absorption variations in Seyfert 
galaxies: 1) changes in the ionization fractions due to ionizing continuum 
variations, and 2) changes in the total column of gas due, for example, to bulk 
motion of the gas across the line of sight (Crenshaw et al. 1999, and references 
therein). The former can be identified from a correlation between continuum and 
absorption variations.
The best case for correlated variations has been established for the neutral 
hydrogen absorption and low-ionization lines of NGC~4151 on time scales of days 
(Kriss et al. 1996; Espey et al. 1998).

For the same Seyfert galaxy, we find strong evidence of variability in the 
ionization state of the absorber on
longer time scales in the C~IV region, where we have multiple GHRS and STIS 
observations at high spectral resolutions. The variation is dramatic in the 
sense that low-ionization lines, such as those from Si~II and Fe~II, are weak or 
undetectable when the continuum in NGC~4151 is in a high state, but are strong 
when the continuum 
drops to a low state (corresponding to $\leq$ 1 x 10$^{-13}$ ergs s$^{-1}$ 
cm$^{-2}$ \AA$^{-1}$ at 1450 \AA). Looking back at the {\it IUE} observations, 
low states defined in this manner occured on several occasions over 1 -- 2 year 
periods between 1978 and 1990 (Ulrich et al. 1991), and the published {\it IUE} 
spectra show that the low-ionization lines are strong on these occasions.
This impression is supported by the finding of Bromage et al. (1985) that the 
equivalent widths of the low-ionization lines (e.g., Si~II, Mg~II) are 
anti-correlated with continuum flux. Thus, the appearance of low-ionization 
lines when the continuum reaches a low state is a long-term phenomenon, spanning 
at least the last two decades.

\subsection{The transient absorption feature in GHRS spectra}

The absorption feature at 1535 \AA\ that Weymann et al. find in epoch 4 of their 
GHRS spectra and suggest is a ``transient'' C~IV feature at high velocity is 
actually the D component of the fine-structure Si~II$^{*}$ $\lambda$1533.4 line. 
This identification is firm, since the velocity centroid and full-width at 
half-maximum (FWHM) of this feature are essentially identical in the GHRS epoch 
4 and STIS echelle spectra, and match the values determined for the D component 
from other lines.
However, this feature is much stronger at the low continuum level of the STIS 
echelle spectrum (a hint of this feature can also be seen in the STIS 
low-resolution spectrum from 1998). The narrow absorption line in the red wing 
of this feature at 1537.5 \AA\ (see Figure 1) is the E$'$ component of 
Si~II$^{*}$ $\lambda$1533.4 (at $-$210 km s$^{-1}$) which is identified by 
Weymann et al. in Mg~II but not in C~IV (and is also present in Si~II 1526.7). 
Interestingly, the broad E component (at $-$ 255 km 
s$^{-1}$) appears to be weak or missing in the Si II$^{*}$ and metastable Fe~II 
lines, indicating  a lower density for this component.

One puzzling aspect of the Si~II$^{*}$ absorption in the GHRS spectra is that 
it is seen in a relatively high state (GHRS epoch 4), similar to that of our 
1997 STIS spectrum, and not seen in a lower state (GHRS epoch 2).
One possible explanation is that the density of the gas responsible for this 
absorption is low enough that it has not had time to respond to a previous 
(unobserved) continuum change. This requires the recombination time scale to be 
greater than or equal to the time scale for large-amplitude continuum 
variations, which is $\geq$2 days (Crenshaw et al. 1996). However, the density 
of the gas would have to be n$_{e}$ $\leq$ $[$$\alpha$$_{rec}$ t$_{rec}$]$^{-1}$ 
$\leq$ $[$(10$^{-12}$ cm$^{-3}$ s$^{-1}$)(1.73 x 10$^{5}$ s]$^{-1}$  $\leq$ 5.8 
x 10$^{6}$ cm $^{-3}$, which is much lower than the derived value of n$_{e}$ 
$\approx$ 10$^{9}$ cm $^{-3}$ for the D component based on the C~III$^{*}$ 
metastable line.
Another possible explanation is that the EUV and 
UV continuum fluxes are not well correlated, and therefore the observed UV 
continuum is not a good indicator of the ionizing flux (at photon energies 
$\geq$ 16.4 eV for Si~II to Si~III). We consider this possibility 
unlikely as well, since UV, EUV, and X-ray continuum variations appear to be 
correlated in this object (Edelson et al. 1996) and a number of other Seyfert 
galaxies (e.g., Chiang et al. 2000, Nandra et al. 2000). Thus, this issue 
remains open at present.

\subsection{The C~IV satellite lines}

A number of {\it IUE} observations have shown narrow emission features in the 
wings of the broad C~IV emission line that could not be attributed to typical 
Seyfert emission lines at the redshift of NGC~4151 (Ulrich et al. 1985; Clavel 
et al. 1987; Ulrich 1996). These lines were relatively narrow (FWHM $\approx$ 7 
-- 16 \AA), located on either side of the C~IV peak, and designated L1 and L2 at 
the observed positions of 1518 \AA\, and 1594 \AA\ (Clavel et al. also 
note a subcomponent L2$'$ located at 1576 \AA). These features 
have been identified as C~IV ``satellite lines'' at relatively large radial 
velocities, and it was suggested that they may arise from a two-sided jet 
(Ulrich et al. 1985). Interestingly, the satellite lines were seen primarily in 
continuum low states, and were not detected in {\it HST} spectra (prior to our 
observations), which were obtained at moderate to bright states.

To investigate this phenomenon, we retrieved an {\it IUE} SWP spectrum obtained 
on 1983 November 19 from the {\it IUE} archives. This spectrum was obtained when 
the continuum flux was low, and shows the satellite lines prominently (Clavel et 
al. 1987). Figure 3 shows the {\it IUE} spectrum in the C~IV region, along with 
the STIS echelle spectrum. To compare the two spectra, we resampled the STIS 
spectrum to the {\it IUE} bin size (1.67 \AA\ per bin) and smoothed it to the 
{\it IUE} resolution ($\sim$ 5 \AA). The binned STIS spectrum is plotted as a 
dashed line in Figure 3.

Figure 3 demonstrates that the satellite lines (L1, L2$'$, and L2) in the {\it 
IUE} spectra are also present in the binned STIS spectrum. In the original STIS 
spectrum these lines are seen to be regions of relatively unabsorbed 
broad-emission plus continuum in the wings of C~IV, surrounded by broad 
low-ionization absorption lines (primarily from Si~II and Fe~II). The prominence 
of the ``satellite lines'' at low continuum states is therefore explained by the 
increased strength of the surrounding low-ionization absorption lines when the 
continuum flux is low. 

\section{Conclusions}

Our STIS spectra show that the broad low-ionization absorption lines appear in 
NGC~4151 when the UV continuum flux is low, and are therefore very sensitive to 
changes in the ionizing flux. This is apparently a long term phenomenon, since 
evidence for an anti-correlation between the equivalent widths of the low 
ionization lines and the UV continuum fluxes has been established for {\it IUE} 
observations dating back to 1978 (Bromage et al. 1985). High-ionization lines 
such as C~IV are apparently close to saturation at both high 
and low states (with evidence for a residual flux in the troughs of these lines 
which is likely due to scattered light near the nucleus, see Kraemer et al. 
2000).

The STIS echelle observations reveal a huge number of previously unidentified 
broad (FWHM $\approx$ 400 km s$^{-1}$) absorption features in the low-state 
spectrum of NGC~4151. We identify most of these features as Fe~II lines from 
multiplets that arise from metastable levels as high as 4.1 eV above the ground 
state. Although this type of absorption is rare, it has been seen in a few 
active galaxies and quasars with higher luminosities (e.g., Mrk 231, Smith et 
al. 1995; Q0059$-$2735, Wampler et al. 1995; Arp 102B, Halpern et al. 1996), and 
is indicative of low-ionization gas with at least moderate electron densities 
(n$_{e}$ $>$ 10$^{6}$ cm$^{-3}$).
 
The variability of the low-ionization lines helps to explain a couple of 
puzzling aspects of the UV spectrum of NGC~4151. The transient feature seen in 
the blue wing of the C~IV emission line in GHRS spectra is actually the D 
component of the Si~II$^{*}$ $\lambda$1533.4 fine-structure line. The C~IV 
``satellite lines'' seen in {\it IUE} spectra are regions of unabsorbed 
continuum plus broad C~IV emission that become prominent when the low-ionization 
absorption appears at low continuum levels. The basic nature of this absorption, 
however, is not well understood; future efforts will concentrate on the 
kinematics and physical conditions in the UV absorbers in NGC 4151.

\acknowledgments
We thank Richard Mushotzky and Fred Bruhweiler for helpful discussions and 
suggestions.
This work was supported by NASA Guaranteed Time Observer funding to the STIS 
Science Team under NASA grant NAG 5-4103.

\clearpage

\figcaption[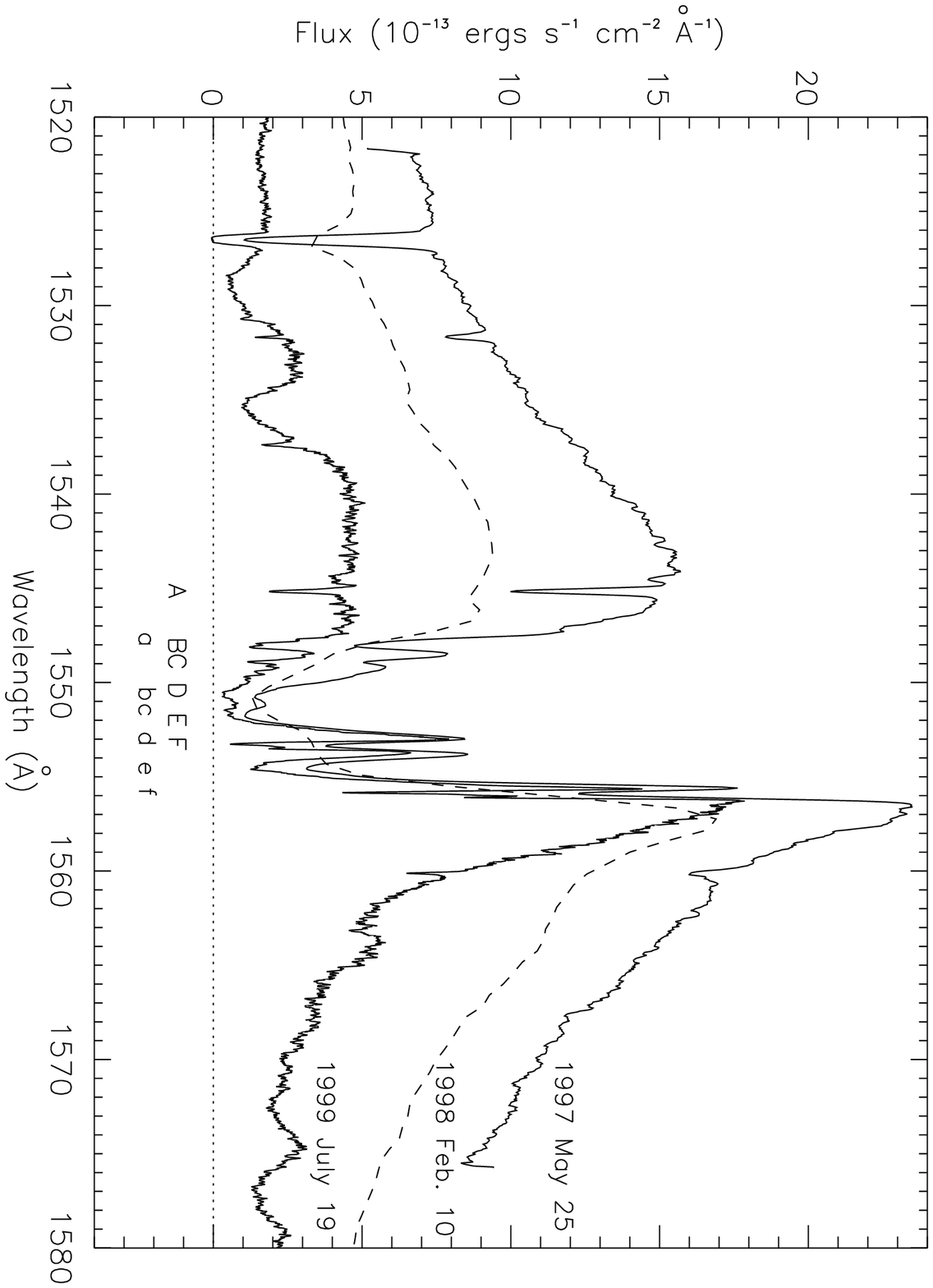]{STIS spectra of the nucleus of NGC~4151 in the C~IV region 
at three epochs. The low-resolution spectrum from 1998 is plotted as a dashed 
line. The positions of the major kinematic components of absorption from Weymann 
et al. (1997) are given in capital letters for the C~IV $\lambda$1548.2 line and 
small letters for the C~IV $\lambda$1550.8 line.}

\figcaption[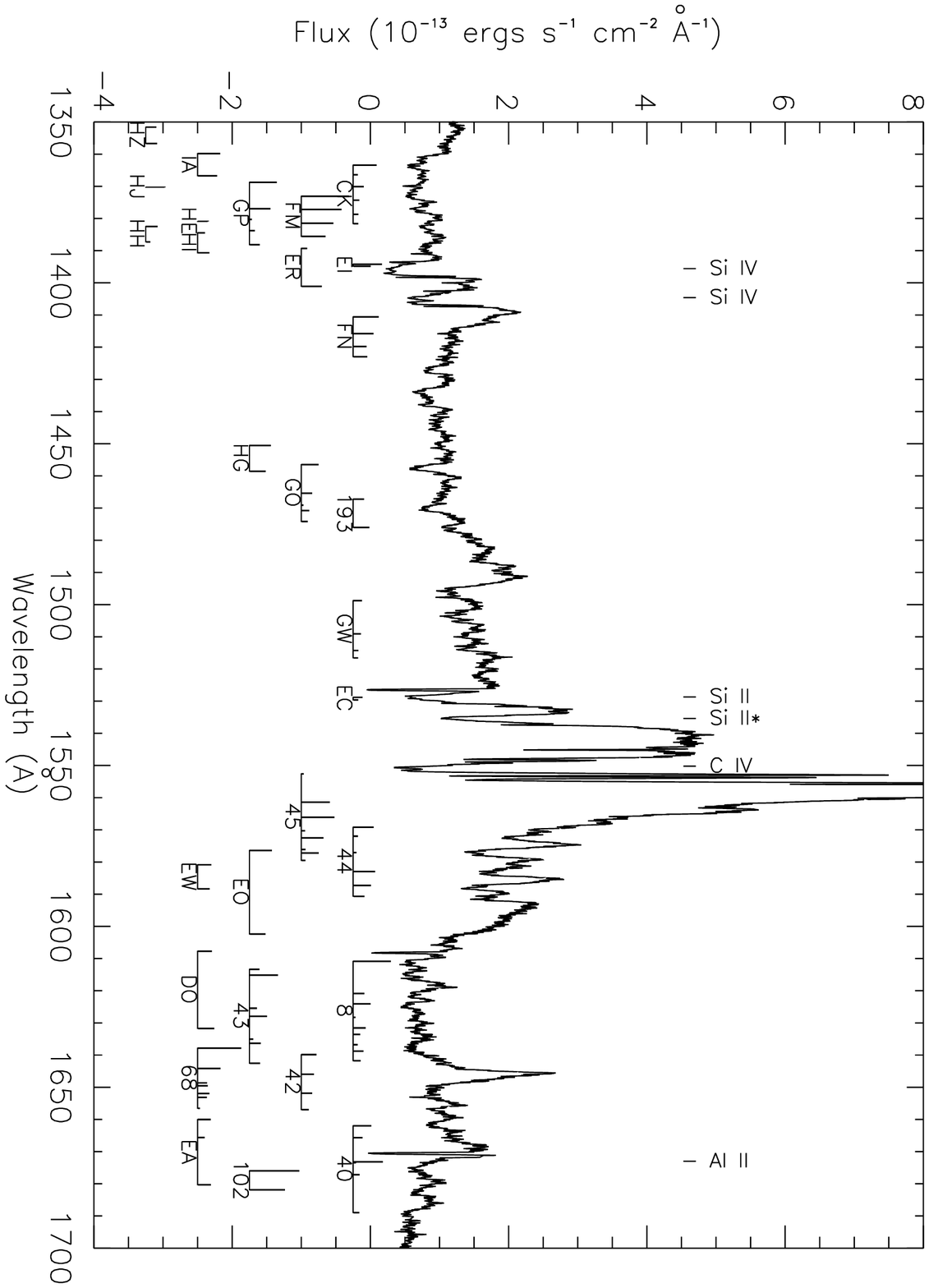]{Identification of the broad absorption lines in the 1350 -- 
1700 \AA\ region; vertical lines represent the locations of the D component (at 
$-$560 km s$^{-1}$, with respect to systemic). Fe~II multiplets in this region 
are plotted below the spectrum, with the length of the vertical lines 
representing the strength of the {\it gf} values; multiplets that have at least 
one line with log ({\it gf}) $>$ $-$0.7 from the lists of Silvis \& Bruhweiler 
(2000) are plotted.}

\figcaption[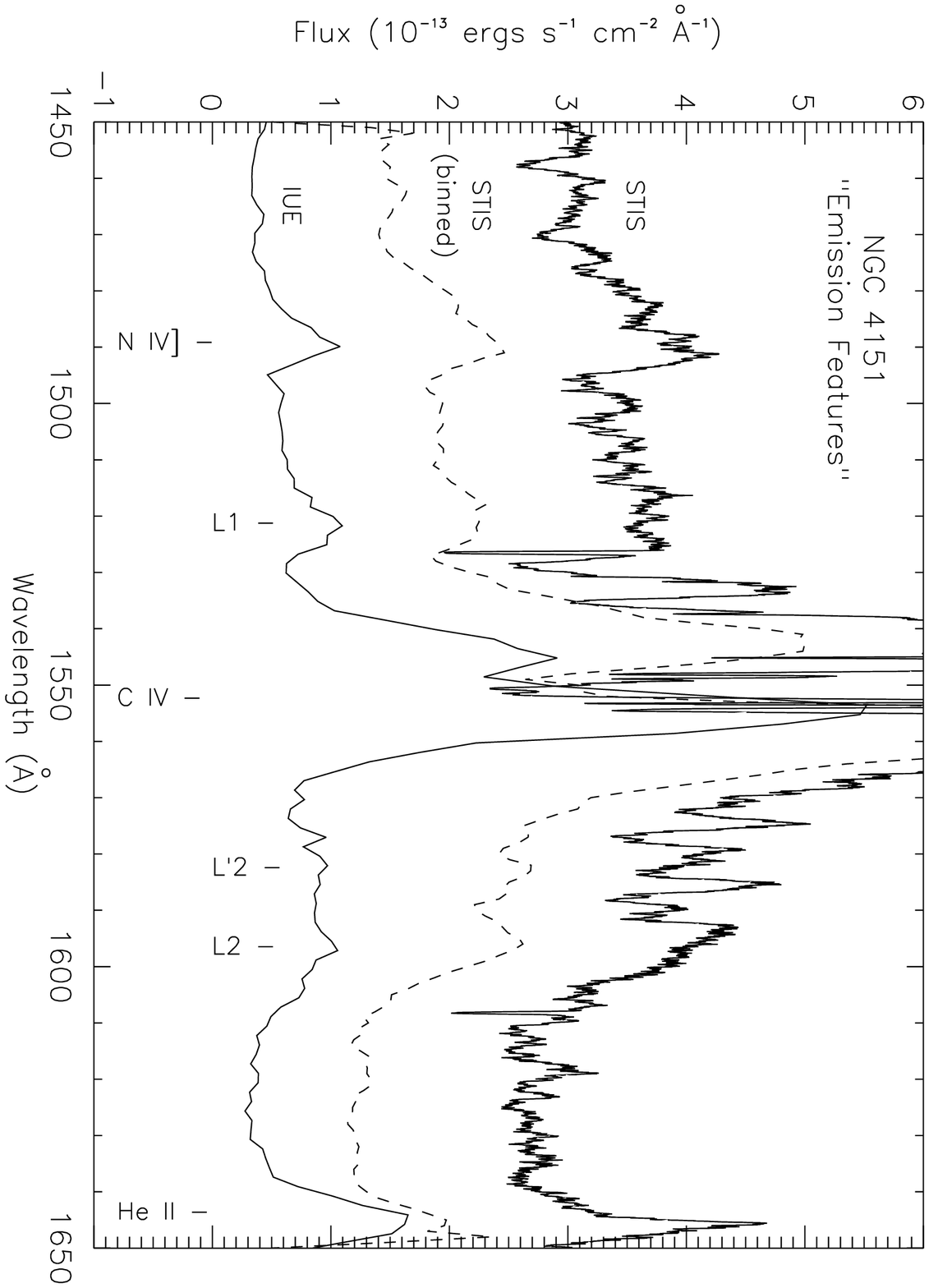]{Comparison of STIS and IUE spectra for NGC~4151 at low 
continuum states. The original STIS spectrum is offset in flux by 2.0 x 
10$^{-13}$ ergs s$^{-1}$ cm$^{-2}$ \AA$^{-1}$ and the binned STIS spectrum is 
offset by 0.5 x 10$^{-13}$ ergs s$^{-1}$ cm$^{-2}$ \AA$^{-1}$. The locations of 
emission lines and the ``satellite lines'' are given.}

\clearpage
\begin{deluxetable}{lccccr}
\tablecolumns{6}
\footnotesize
\tablecaption{STIS UV Observations of the Nucleus of NGC~4151 \label{tbl-1}}
\tablewidth{0pt}
\tablehead{
\colhead{Date (UT)} & \colhead{Grating} & \colhead{Aperture}  &
\colhead{Exp. Time} & \colhead{Coverage} & \colhead{Resolution}\\
\colhead{} &\colhead{} &\colhead{} &
\colhead{(sec)} &\colhead{(\AA)} &\colhead{(km s$^{-1}$)}
}
\startdata
\multicolumn{6}{c}{Echelle Spectra} \\
\tableline \\
1999 July 19/20 &E140M &0\arcsecpoint2 x 0\arcsecpoint2 &5546 &1146 -- 1710 &7\\
                &E230M &0\arcsecpoint2 x 0\arcsecpoint2 &2632 &1614 -- 2360 
&10\\
                &E230M &0\arcsecpoint2 x 0\arcsecpoint2 &2304 &2278 -- 3092  
&10\\
\tableline
\multicolumn{6}{c}{} \\
\multicolumn{6}{c}{Previous STIS Spectra}\\
\tableline
1997 May 25      &G140M  &25$''$ x 25$''$         &2369 &1522 -- 1576  &20\\
1998 January 8   &G230LB &25$''$ x 0\arcsecpoint2 &2160 &1680 -- 3060 &300\\
1998 February 10 &G140L  &25$''$ x 0\arcsecpoint2 &4500 &1150 -- 1730 &240\\

\enddata
\end{deluxetable}


\clearpage
\vskip3.0in
\begin{figure}
\plotone{fig1.ps}
\\Fig.~1.
\end{figure}

\clearpage
\vskip3.0in
\begin{figure}
\plotone{fig2.ps}
\\Fig.~2.
\end{figure}

\clearpage
\vskip3.0in
\begin{figure}
\plotone{fig3.ps}
\\Fig.~3.
\end{figure}

\end{document}